\documentclass{article}
\usepackage{amsmath,amssymb,epsfig}
\usepackage{graphicx}

\begin{document}

\title{Ambiguities of theoretical parameters and CP/T violation in
     neutrino factories}

\author{%
   Masafumi Koike%
     \footnote{e-mail address: {\tt koike@icrr.u-tokyo.ac.jp}}\\%
     {\footnotesize \it%
     Institute for Cosmic Ray Research, University of Tokyo,}\\
     {\footnotesize \it%
     Kashiwa-no-ha 5-1-5, Kashiwa, Chiba 277-8582, Japan%
     }
   \\
   Toshihiko Ota%
     \footnote{e-mail address: {\tt toshi@higgs.phys.kyushu-u.ac.jp}}\\%
     {\footnotesize \it%
     Department of Physics, Kyushu University, Fukuoka 812-8581, Japan
     }%
   \\
   and
   \\
   Joe Sato%
     \footnote{e-mail address: {\tt joe@rc.kyushu-u.ac.jp}}\\%
     {\footnotesize \it%
     Research Center for Higher Education, Kyushu University,}\\
     {\footnotesize \it%
     Ropponmatsu, Chuo-ku, Fukuoka 810-8560, Japan%
   }%
}


\maketitle

\begin{abstract}
  We study the optimal setup for observation of the CP asymmetry in
neutrino factory experiments --- the baseline length, the muon energy
and the analysis method.
First, we point out that the statistical quantity which has been used in
previous works doesn't represent the CP asymmetry.
Then we propose the more suitable quantity, $\equiv \chi^{2}_{2} $,
which is sensitive to the CP asymmetry.  We investigate the behavior of
$ \chi^{2}_{2} $ with ambiguities of the theoretical parameters.  The
fake CP asymmetry due to the matter effect increases with the baseline
length and hence
 the error in the estimation of the fake CP asymmetry grows
 with the baseline length due to the ambiguities of the theoretical
parameters. Namely, we lose the sensitivity to the genuine CP-violation
 effect in longer baseline.
 \end{abstract}

\section{Introduction}

The observation of the atmospheric neutrino anomaly by Super-Kamiokande
\cite{AtmSK} provided us with convincing evidence that neutrinos
have non-vanishing masses. There is another indication of neutrino
masses and mixings by the solar
neutrino deficit \cite{Ga1,Ga2,Kam,Cl,SolSK}.

Assuming three generations of the
leptons, we denote the lepton mixing matrix, which relates the flavor
eigenstates ($\alpha = {\rm e},\mu,\tau$) with the mass eigenstates with
mass $m_{i} (i=1,2,3)$, by
\begin{eqnarray}
    U_{\alpha i}
    =
    \begin{pmatrix}
	c_{13} c_{12}
	&
	c_{13} s_{12}
	&
	s_{13}
	\\
	- c_{23} s_{12} - s_{23} s_{13} c_{12} {\rm e}^{{\rm i} \delta}
	&
	c_{23} c_{12} - s_{23} s_{13} s_{12} {\rm e}^{{\rm i} \delta}
	&
	s_{23} c_{13} {\rm e}^{{\rm i} \delta}
	\\
	s_{23} s_{12} - c_{23} s_{13} c_{12} {\rm e}^{{\rm i} \delta}
	&
	-s_{23} c_{12} - c_{23} s_{13} s_{12} {\rm e}^{{\rm i} \delta}
	&
	c_{23} c_{13} {\rm e}^{{\rm i} \delta}
    \end{pmatrix}_{\alpha i},
\end{eqnarray}
where $c_{ij} (s_{ij})$ is the abbreviation of $\cos\theta_{ij}
(\sin\theta_{ij})$. Then
the atmospheric neutrino anomaly gives an allowed region for
$\sin\theta_{23}$ and the larger mass square difference ($\equiv\delta
m^2_{31}$). The solar neutrino deficit provides allowed regions for
$\sin\theta_{12}$
($\equiv \delta m^2_{21}$).

On the other hand, there is only an excluded region for
$\sin\theta_{13}$ from reactor experiments~\cite{Chooz}. 
Furthermore there is no constraint on the CP violating phase $\delta$. 
The idea of neutrino factories with muon storage rings were proposed
\cite{Geer}
to determine these mixing parameters%
, and attracted the interest of many physicists
\cite{BGW,Golden,BGRW,BGRW2,GH,DFLR,FLPR,NuFact,BCR,KS,Yasuda}. 

However we have some questions concerning the previous analyses of the
CP-violation effect. In many analysis, the muon energy of $ E_{\mu}$ is
assumed to be rather high.  This seems strange since CP/T violation
arises as a three generation effect \cite{KM,KS,three,kos1}. Indeed we
can derive very naively that $E_{\mu} \sim 30 {\rm GeV}$,  lower
by factor 2, is the most efficient for $L = 3000$ km \cite{KS} while
$E_{\mu} \sim 50 {\rm GeV}$ is often assumed. Furthermore the fake
CP-violation effect due to the matter effect\cite{MSW} increases with
baseline length.  The ambiguity in the estimate for the
fake CP violation increases with baseline length.  Taking into
account this ambiguity in the analysis, the
sensitivity to CP violation will be decreased as baseline length
increases.  It is unlikely that we can observe the CP-violation
effect with such a long baseline. We discuss these problems\cite{kos1}.


\section{Statistical quantity}
\label{subsect:proper-chi2}

As an experimental setup, we consider that $N_{\mu}$ muons decay at a muon
ring.  The neutrinos extracted from the ring are detected at a detector
if $E_{\nu}$ is larger than a threshold energy $E_{\rm th}$.  The
detector has mass $M_{\rm detector}$ and contains $N_{\rm target}$
target atoms.  We assume that the neutrino-nucleon cross section $\sigma$
is proportional to neutrino energy as
\begin{equation}
    \sigma = \sigma_{0} E_{\nu},
    \label{eq:sigma}
\end{equation}
The expected number of appearance events in the energy bin $E_{j-1} <
E_{\nu} < E_{j}$ ($j = 1, 2, \ldots, n$) is then given by

\begin{equation}
    N_{j} (\nu_{\alpha} \rightarrow \nu_{\beta}; \delta)
    \equiv
    \frac{ N_{\mu} N_{\rm target} \sigma_{0} }{ \pi m_{\mu}^{2} }
    \frac{ E_{\mu}^{2} }{ L^{2} } 
    \int_{ E_{j-1} }^{ E_j }
    E_{\nu}
    f_{\nu_{\alpha}} (E_{\nu})
    P(\nu_{\alpha} \rightarrow \nu_{\beta}; \delta)
    \frac{ {\rm d} E_{\nu} }{ E_{\mu} },
    \label{eq:app-event-number}
\end{equation}
where $m_{\mu}$ is the muon mass. 

To estimate the sensitivity for the CP-violation effect
the following statistics is usually used:
\begin{eqnarray}
    \chi_{1}^{2} (\delta_{0})
    & \equiv &
    \sum_{j=1}^{n}
    \frac
    {[ N_{j}(\delta) - N_{j}(\delta_{0}) ]^{2}}
    {N_{j}(\delta)}
    +
    \sum_{j=1}^{n}
    \frac
    {[ \bar N_{j}(\delta) - \bar N_{j}(\delta_{0}) ]^{2}}
    {\bar N_{j}(\delta)}
    \label{eq:chi1-0-def}
\end{eqnarray}
$n$ is the number of bins. Since the CP violation is absent if $\sin
\delta = 0$, namely $\delta = 0$ or $\delta = \pi$, we need to check
that $N_{j}(\delta)$ is different from $N_{j}(\delta_0)$ with
$\delta_{0} \in \{0, \pi \}$ to insist that CP violation is present.

We can claim that $N_{j}(\delta)$ is different
from $N_{j}(\delta_0)$ at 90\% confidence level, if
\begin{eqnarray}
    \chi^{2}_{1}
    &\equiv&
    \min(\chi^{2}_{1}(0), \chi^{2}_{1}(\pi))
> \chi^{2}_{90\%}(n)
    \label{eq:chi1-def}
\end{eqnarray}
holds. Here $\chi^2_{90\%} (n)$ is the $\chi^2$ value with $n$ degrees
of freedom at 90\% confidence level. 

To see the behavior of $\chi_1^2$,
we make use of high energy approximation which is valid for $E_{\nu}
\gtrsim (\delta m^{2}_{31} L) / 4$:
\begin{equation}
    \chi_{1}^{2}(\delta_{0})
    \propto
    E_{\mu}
    \frac{ J_{/\delta}^{2} }{A}
    \left\{
    (\cos \delta \mp 1)
    \left[
    1 -
    \frac{1}{3}
    \left(
    \frac{a(L) L}{4 E_{\nu}^{\rm peak}}
    \right)^{2}
    \right]
    \right\}^{2}
    \label{eq:chi1-highE}
\end{equation}
Here $E_{\nu}^{\rm peak}$ is the neutrino energy which gives the maximum
valu of the initial neutrino flux $f_{\nu_\alpha}$ and $a(L)$ is the
effective mass square due to matter effect calculated using Preliminary
Reference Earth Model(PREM)\cite{PREM,Ota}.  We find that $\chi_{1}^{2}$
is an increasing function of $E_{\mu}$.  Thus we can obtain arbitrary
large $\chi_{1}^{2}$, and we can seemingly achieve arbitrary high
sensitivity to search for the CP-violation effect, by increasing muon
energy.  Thus the higher energy appears to be preferable to observe the
CP-violation effect as long as we employ $\chi_{1}^{2}$. It is
important, however, to note that $\chi_{1}^{2}$ has nothing to do with
the imaginary part of the mixing matrix in high energy limit.  The CP
violation is brought about by the only imaginary part of the mixing
matrix, which is proportional to $\sin \delta$ in our
parameterization. $\chi_1^2$ is relevant with CP violation
through unitarity\cite{three}.

Therefore we need to consider a statistical quantity
which is sensitive to the imaginary part of the lepton mixings.
As such a statistics we consider the following quantity:\cite{GH,DFLR}
\begin{equation}
    \chi_{2}^{2} (\delta_{0})
    \equiv
    \sum_{j=1}^{n}
    \frac
    {\left[
    \Delta N_{j}(\delta) - \Delta N_{j}(\delta_{0}) \right]^{2}
    }{
    N_{j}(\delta) + \bar N_{j}(\delta)
    }
    \label{eq:chi2-0-def}
\end{equation}
Here $\Delta
N_{j}(\delta) \equiv N_{j}(\delta) - \bar N_{j}(\delta) $.
It is required
\begin{eqnarray}
    \chi^{2}_{2} &\equiv& \min (\chi^{2}_{2}(0), \chi^{2}_{2}(\pi))
> \chi^{2}_{90\%}(n)
\label{eq:chi2-def}
\end{eqnarray}
to claim that CP violation effect is observed.

In the high energy limit
\begin{equation}
    \chi_{2}^{2}(\delta_{0})
    \propto
    \frac{L^{2}}{E_{\mu}}
    \frac{ J_{/\delta}^2 }{A}
    \left\{
    \sin\delta +
    \frac{1}{3}
    \frac{ a(L) L }{4 E_{\nu}^{\rm peak}}
    (2 \cos 2\theta_{13} - 1) (\cos \delta \mp 1)
    \right\}^{2}
    \label{eq:chi2-highE}
\end{equation}
($-$ for $\delta_{0} = 0$ and $+$ for $\delta_{0} = \pi$), where

\begin{eqnarray}
    J_{/\delta}
    &\equiv&
    \frac{\delta m_{21}^{2}}{\delta m_{31}^{2}}
    \sin 2\theta_{12} \sin 2\theta_{23}
    \sin 2\theta_{13} \cos\theta_{13},
    \label{eq:J/delta-def}
    \\
    A
    &\equiv&
    \sin^2 \theta_{23} \sin^2 2\theta_{13}.
    \label{eq:Adef}
\end{eqnarray}

To see CP-violation effect is to measure $J_{/\delta}\sin\delta$\cite{AJ,AKS}.
In this respect $\chi_{2}^{2}$ gives a good standard to observe CP
violation.


\section{Feasibility of CP violation search in presence of the 
ambiguities of the parameters}
\label{sect:CP-violation}

In this section we study the asymmetry with $\chi_{2}^{2}$.
The values of all theoretical parameters will have ambiguities in
practice, and hence we cannot estimate $\Delta N_{j}(\delta_{0})$
precisely.  The genuine CP-violation effect will be absorbed into the
ambiguity of $\Delta N(\delta_{0})$ if the ambiguity of $\Delta
N(\delta_{0})$ is large.  Therefore we must examine whether
the CP-violation effect can be absorbed in the ambiguities of the
parameters.

Suppose that we use the parameters $\tilde x_{i} \equiv \{ \tilde
\theta_{kl}, \delta \tilde m^{2}_{kl}, \tilde a(L) \}$, which are
different from the true values $x_{i} \equiv \{ \theta_{kl}, \delta
m^{2}_{kl}$, $a(L) \}$, to calculate $N_{j}(\delta_{0})$ and $\bar
N_{j}(\delta_{0})$.  We will estimate the fake CP violation due to the
matter effect as
\begin{eqnarray}
    \Delta \tilde N_{j}(\delta_{0})
    &=&
    \tilde N_{j}(\delta_{0})
    -
    \tilde{\bar N_{j}}(\delta_{0}),
    \label{eq:Delta-tilde-Nj}
%
%
\end{eqnarray}
are evaluated from eqs.(\ref{eq:app-event-number}).  We then obtain
\begin{equation}
    \tilde \chi^{2}_{2} (\delta_{0})
    \equiv
    \sum_{j=1}^{n}
    \frac
    { [ \Delta N_{j}(\delta) - \Delta \tilde N_{j}(\delta_{0}) ]^{2} }
    { N_{j}(\delta) + \bar N_{j}(\delta) }
    \label{eq:chi2tilde-0-def}
\end{equation}
%
%
instead of $\chi^{2}_{2}(\delta_{0})$. 
The observed
asymmetry $\Delta N_{j}(\delta)$ consists of the genuine CP-violation
effect and the fake one due to the matter effect.  We have to subtract
the matter effect, but we cannot estimate precisely the fake CP
violation $\Delta \tilde N_{j}(\delta_{0})$ due to the ambiguities of
the parameters.  In such a case the sensitivity to CP-violation search
gets worse once the ambiguities of the parameters are taken into
account, since it is always possible to take $\Delta \tilde
N_{j}(\delta_{0})$ to satisfy
\begin{equation}
    \left|
    \Delta N_{j}(\delta) - \Delta \tilde N_{j}( \delta_{0})
    \right|
    \le
    \left|
    \Delta N_{j}(\delta) - \Delta N_{j}(\delta_{0})
    \right|,
    \label{eq:worse-estimate}
\end{equation}
or equivalently
\begin{equation}
    \tilde \chi_{2}^{2} \leq \chi^{2}_{2},
    \label{eq:worse-chi2}
\end{equation}
by adjusting $\tilde x_{i}$'s.  We can further argue that we lose
more sensitivity as the baseline length gets longer.
Let us illustrate the outline described above in detail.  The CP
asymmetry of probabilities

\begin{equation}
    A(\{ x_{i} \}, \delta)
    \equiv
    P(\nu_{\alpha} \rightarrow \nu_{\beta}; \{ x_{i} \}, \delta)
    -
    P(\bar \nu_{\alpha} \rightarrow \bar \nu_{\beta};
      \{ x_{i} \}, \delta)
    \label{eq:Asym-def}
\end{equation}
consists of the genuine CP asymmetry $A_{\rm CPV}(\{ x_{i} \}, \delta)$
and the fake one $A_{\rm CPM}(\{ x_{i} \}, \delta)$, so that
\begin{equation}
    A(\{ x_{i} \}, \delta)
    =
    A_{\rm CPV}(\{ x_{i} \},\delta) + A_{\rm CPM}(\{ x_{i} \}, \delta).
    \label{eq:A-delta}
\end{equation}
We need to subtract $A(\{ x_{i} \}, \delta_{0})$ from $A(\{ x_{i} \},
\delta)$, but instead we subtract $A(\{ \tilde x_{i} \}, \delta_{0})$
due to the ambiguities of the parameters and obtain
\begin{eqnarray}
    \tilde A_{\rm CPV}(\delta)
    &\equiv&
    A_{\rm CPV}(\{ x_{i} \},\delta) +
    A_{\rm CPM}(\{ x_{i} \}, \delta) -
    A(\{ \tilde x_{i} \}, \delta_{0}).
    \label{eq:Atilde}
\end{eqnarray}
Here $A_{\rm CPV}$ and $A_{\rm CPM}$ can be estimated using high
energy approximation as
\begin{eqnarray}
    A_{\rm CPM}(\{ x_{i} \}, \delta)
    &\simeq&
    \frac{1}{3}
    \left[
    2
    \sin^2 \theta_{23}
    \sin^2 2\theta_{13}
    \cos 2\theta_{13}
\right.
       \nonumber \\
& +&
\left.    (2 \cos 2\theta_{13} - 1)
    J_{/\delta}
    \cos\delta
    \right]
    \frac{a(L) L}{4E_\nu}
    \left( \frac{\delta m_{31}^2 L}{4E_\nu} \right)^3
    \label{eq:fakeCP}\\
    A_{\rm CPV}(\{ x_{i} \}, \delta)
    &=&
    \left( \frac{\delta m_{31}^2 L}{4 E_\nu}\right )^3
    J_{/\delta}
    \sin\delta.
    \label{eq:trueCP}
\end{eqnarray}
The factor 
\begin{equation}
    \frac{2}{3} \sin^{2} \theta_{23} \sin^{2} 2\theta_{13} \cos
    2\theta_{13} \frac{a(L) L}{4 E_\nu}
    \label{eq:fakeCPdominant}
\end{equation}
in eq.(\ref{eq:fakeCP}) is expected to be much larger than
$J_{/\delta}$ in eq.(\ref{eq:trueCP}) with a long baseline
. Thus the ambiguity of the fake CP-violation effect, $A_{\rm CPM}(\{
x_{i} \}, \delta) - A(\{ \tilde x_{i} \}, \delta_{0})$, can absorb the
genuine CP-violation effect $A_{\rm CPV}$, so that $\tilde A_{\rm
CPV}$ , or equivalently $\tilde \chi_{2}^{2}$, becomes significantly
small.  The condition to observe CP-violation effect in 90\%
confidence level, say, is again given by
\begin{eqnarray}
    X_{2}^{2} &\equiv& \min_{ \{ \tilde x_{i} \} } \tilde{\chi}_{2}^{2}
> \chi^{2}_{\rm 90\%}(n).
    \label{eq:tilde-chi2-cond}
\end{eqnarray}

\begin{figure}
    \unitlength=1cm
    \begin{picture}(15,7)
\includegraphics[width=8cm]{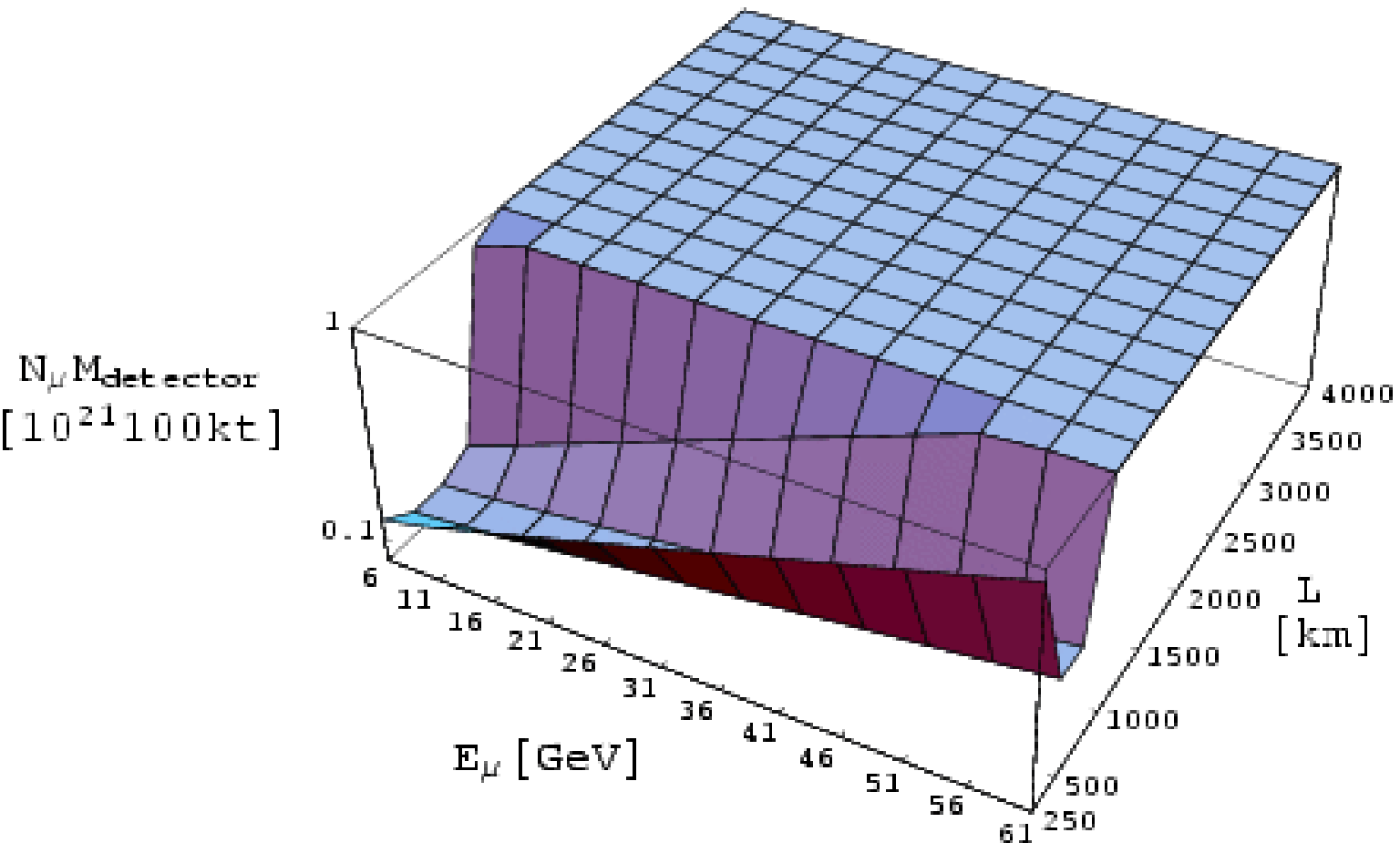}
\includegraphics[width=6cm]{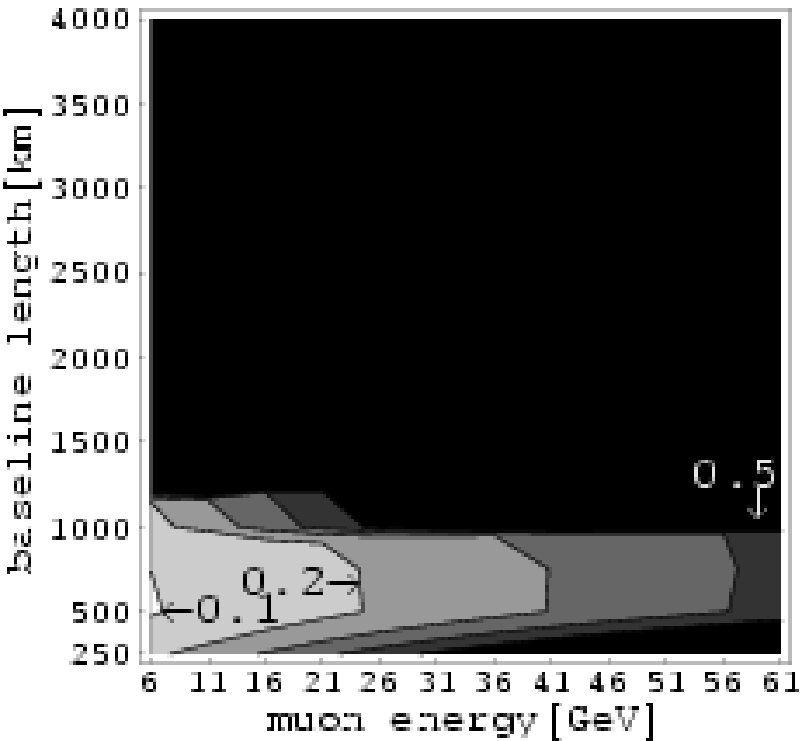}
    \end{picture}
    \caption{Necessary value of $N_{\mu} M_{\rm detector}$ to observe
    the CP-violation effect as a function of muon energy and baseline
    length, for $\delta = \pi / 2$ and $E_{\rm th} = 1 {\rm GeV}$. Here
    ($\sin\theta_{13}$,$\sin\theta_{23}$,$\sin\theta_{12}$,
 $\delta m^{2}_{31}$,$\delta m^{2}_{21}$) = (0.1, 1/$\sqrt{2}$, 0.5,
    3$\times 10^{-3}$eV$^2$, $10^{-4}$eV$^2$) and $a(L)$ is calculated
    using PREM.
  The ambiguities of the theoretical
    parameters are assumed to be 10 \%. 
 Hence these graphs are
    obtained using not $\chi_{2}^{2}$ but $ X_{2}^{2}$.
    The sensitivity to the genuine CP asymmetry is lost in long
    baseline region such as $L \gtrsim 1250 {\rm km}$ as we estimate in
    eq.(\ref{eq:L-condition2}).}
  \label{fig:cp-ambiguity}
\end{figure}

We present in Figs.\ref{fig:cp-ambiguity} the required value of $N_{\mu}
M_{\rm detector}$ obtained from eq.(\ref{eq:tilde-chi2-cond}) to observe
the CP-violation effect in 90\% confidence level. 
All the parameters are assumed to have
ambiguities of 10 \%.
We find that we cannot observe the genuine
CP-violation effect when $L$ is larger than 1000 km.  We can
qualitatively understand it by eqs.(\ref{eq:fakeCP}) and
(\ref{eq:trueCP}).  It is seen that

\begin{equation}
    \frac{A_{\rm CPV}}{A_{\rm CPM}}
    =
    3
    \frac
    {J_{/\delta} \sin\delta}
    {2 \sin^2 \theta_{23} \sin^2 2\theta_{13} \cos 2\theta_{13}
    +
    (2 \cos 2\theta_{13} - 1) J_{/\delta} \cos\delta}
    \frac{4 E}{a(L) L}
    \label{eq:relativeP}
\end{equation}
is a decreasing function of $L$, which means that the sensitivity to
the CP violation is lost as the baseline length gets larger.  The
condition on $L$ is roughly estimated by $A_{\rm CPV} / A_{\rm CPM}
\gtrsim 1$, or
\begin{equation}
    L
    \lesssim
    \frac{4 E}{a(L)}
    \frac{3 J_{/\delta} \sin\delta}
    {2 \sin^2\theta_{23} \sin^2 2\theta_{13} \cos 2\theta_{13}
    +
    (2 \cos 2\theta_{13} - 1)
    J_{/\delta} \cos \delta
    }.
    \label{eq:L-condition}
\end{equation}
For the parameters used in Fig.\ref{fig:cp-ambiguity},
\begin{equation}
    L \lesssim 1250 {\rm  km}.
    \label{eq:L-condition2}
\end{equation}
%

%

\section{ Summary and Discussion}

We discussed the optimum experimental setup and the optimum analysis
to see the CP violation effect.

We examined how to analyze the data of experiments to confirm the
naive estimation.  We studied with two statistical quantities,
$\chi^2_1$ (eq.(\ref{eq:chi1-def})) and $\chi_2^2$
(eq.(\ref{eq:chi2-def})). Usually $\chi^2_1$ is used in analyses of
neutrino factories. 
We can test by this whether the data can be
explained by the hypothetical data calculated assuming no CP-violation
effect.  
We saw, however, that this quantity is sensitive has information for
mainly the CP
conserved part of the oscillation probability in high energy region.  
Hence we concluded that
it is difficult to measure the CP violation by using this quantity.  
On the other hand, we can test
with $\chi^2_2$ whether the asymmetry of oscillation probabilities of
neutrinos and antineutrinos exists.  We have seen that $\chi^2_2$ is
sensitive to the CP violating part of the oscillation probability, and
thus it is suitable quantity to measure the CP violation.

Then we investigated the influence of the ambiguities of the
theoretical parameters on $\chi^2_2$.  Since the matter effect causes
the difference of the oscillation probabilities between neutrinos and
antineutrinos, we have to estimate the fake asymmetry to search for the CP
violation effect.  However, we will always ``overestimate'' the fake
CP violation due to the ambiguities of the theoretical parameters,
and hence we will always estimate the genuine CP-violation effect too
small.  The matter effect increases as baseline length increases, and
we will lose the sensitivity to the asymmetry due to the genuine
CP-violation effect in longer baseline such as several thousand km.




\end{document}